\def\units#1{\hbox{$\,{\mathrm {#1}}$}}

\def\degrees{\hbox{${}^\circ$}}
\documentstyle[epsfig,12pt]{article}

\begin{document}

%\today
%\hskip 200 pt
%{\bf Draft}
%
%
\begin{center}
{\Large{NUCLEARITE SEARCH \\
\vskip 0.2in
WITH THE MACRO DETECTOR AT GRAN SASSO}}
\end{center}

%\vskip .810 cm

\vskip .7 cm \begin{center}
{\bf The MACRO Collaboration} \\
\nobreak\bigskip\nobreak
\pretolerance=10000
%%%%%%%%%%%%%%%%%%%%%%%%%%%%%%%%%%%%%%%%%%

M.~Ambrosio$^{12}$, 
R.~Antolini$^{7}$, 
C.~Aramo$^{7,n}$,
G.~Auriemma$^{14,a}$, 
A.~Baldini$^{13}$, 
G.~C.~Barbarino$^{12}$, 
B.~C.~Barish$^{4}$, 
G.~Battistoni$^{6,b}$, 
R.~Bellotti$^{1}$, 
C.~Bemporad$^{13}$
% ...
\\
\vspace{1.5 cm}
\footnotesize
1. Dipartimento di Fisica dell'Universit\`a di Bari and INFN, 70126 
Bari,  Italy \\
2. Dipartimento di Fisica dell'Universit\`a di Bologna and INFN, 
40126 Bologna, Italy \\
3. Physics Department, Boston University, Boston, MA 02215, 
USA \\
4. California Institute of Technology, Pasadena, CA 91125, 
USA \\
5. Department of Physics, Drexel University, Philadelphia, 
PA 19104, USA \\
6. Laboratori Nazionali di Frascati dell'INFN, 00044 Frascati (Roma), 
Italy \\
7. Laboratori Nazionali del Gran Sasso dell'INFN, 67010 Assergi 
(L'Aquila),  Italy \\
8. Depts. of Physics and of Astronomy, Indiana University, 
Bloomington, IN 47405, USA \\
9. Dipartimento di Fisica dell'Universit\`a dell'Aquila  and INFN, 
67100 L'Aquila,  Italy \\
10. Dipartimento di Fisica dell'Universit\`a di Lecce and INFN, 
73100 Lecce,  Italy \\
11. Department of Physics, University of Michigan, Ann Arbor, 
MI 48109, USA \\	
12. Dipartimento di Fisica dell'Universit\`a di Napoli and INFN, 
80125 Napoli,  Italy \\	
13. Dipartimento di Fisica dell'Universit\`a di Pisa and INFN, 
56010 Pisa,  Italy \\	
14. Dipartimento di Fisica dell'Universit\`a di Roma ``La Sapienza" and INFN, 
00185 Roma,   Italy \\ 	
15. Physics Department, Texas A\&M University, College Station, 
TX 77843, USA \\	
16. Dipartimento di Fisica Sperimentale dell'Universit\`a di Torino and INFN,
10125 Torino,  Italy \\	
17. L.P.T.P., Faculty of Sciences, University Mohamed I, B.P. 524 Oujda, Morocco \\
18. Dipartimento di Fisica dell'Universit\`a di Roma Tre and INFN Sezione 
Roma III, Roma, Italy \\ 
$a$ Also Universit\`a della Basilicata, 85100 Potenza,  Italy \\
$b$ Also INFN Milano, 20133 Milano, Italy\\
$c$ Also Istituto TESRE/CNR, 40129 Bologna, Italy \\
$d$ Also Scuola Normale Superiore di Pisa, 56010 Pisa, Italy\\
$f$ Also Universit\`a di Trieste and INFN, 34100 Trieste, 
Italy \\
$g$ Also Dipartimento di Energetica, Universit\`a di Roma, 
00185 Roma,  Italy \\
$h$ Also Institute for Nuclear Research, Russian Academy
of Science, 117312 Moscow, Russia \\
$i$ Also Institute for Space Sciences, 76900 Bucharest, Romania \\
$l$ The Colorado College, Colorado Springs, CO 80903, USA\\
$m$ RPD, PINSTECH, P.O. Nilore, Islamabad, Pakistan \\
$n$ Also INFN Catania, 95129 Catania, Italy\\
$o$ Also Department of Physics, Pennsylvania State University, 
University Park, PA 16801, USA\\

\normalsize
\vskip 1.5 cm

{\it Submitted to The European Physical Journal C}
\end{center}
\newpage
{\small
\begin{center}
{ \bf Abstract}
\end{center}
\vspace {-0.6 cm}
\hskip 15 pt

In this paper we present the results of a search for nuclearites in the
penetrating cosmic radiation using the scintillator and track-etch
subdetectors of the MACRO apparatus. The analyses
cover the $ \beta =v/c$ range at the detector depth
(3700 \units{hg/cm^2}) $10^{-5} < \beta < 1$;
for $\beta = 2 \times 10^{-3}$ the flux limit is
$ 2.7\times 10^{-16} \units{cm^{-2}s^{-1}sr^{-1}}$
for an isotropic flux of nuclearites, and twice
this value for a flux of downgoing
nuclearites.

%\newpage
\vspace{7mm}
\section {Introduction}
\vspace{5mm}
\hskip 15 pt

In 1984 Witten formulated the hypothesis \cite{witten}
that ``strange quark matter'' (SQM) composed of comparable
amounts of $u$, $d$ and $s$ quarks  might be the ground state of
hadronic matter.
 ``Bags'' of SQM (also known as ``strangelets'') would be
heavier  than a single $\Lambda\!{}^0$ baryon, but neutral enough
not to be limited in size by Coulomb repulsion.

If particles of SQM were produced in a first-order phase transition in
the early universe, they would be candidates for Dark Matter (DM),
and might be found in the cosmic radiation reaching the Earth.
Particles of SQM in cosmic radiation are commonly known as  ``nuclearites''
\cite{ruhula}. This paper describes an experimental search for nuclearites
with the MACRO detector at Gran Sasso.

The main energy loss mechanism for nuclearites passing through matter is
elastic or quasi-elastic
atomic collisions \cite{ruhula}. The energy loss rate is

\begin{equation}
\frac{dE}{dx} = \sigma \rho v^2,
\end{equation}
where $\sigma$ is the nuclearite cross section,
$v$ its velocity and $\rho$ the mass density of the traversed medium.

For nuclearites with masses $M \geq 8.4 \times 10^{14}$ 
$\units{GeV/c^2}$ ($\simeq 1.5\units{ng}$)
 the cross section may be approximated as \cite{ruhula}:
\begin{equation}
\sigma \simeq \pi \times \left ( \frac{3M}{4 \pi \rho_N} \right ) ^{2/3}
\end{equation}
where $\rho_N$ (the density of strange quark matter) is estimated to be
$\rho_N \simeq 3.5 \times 10^{14}$ \units{g/cm^3}, somewhat larger than that
 of atomic nuclei \cite{chin}.
 For nuclearites of masses $M < 8.4 \times 10^{14}$ \units{GeV/c^2}
the collisions are governed by
their electronic clouds, yielding $\sigma \simeq \pi \times 10^{-16}$ \units
{cm^2}.

An experimental search for nuclearites
    has an acceptance that depends on nuclearite mass,
    since upgoing nuclearites that traverse the Earth before
    detection only occur for sufficiently large nuclearite masses
    ( $\simeq 6 \times 10^{22}$ \units{GeV/c^2} at 
typical galactic velocities \cite{ruhula}).

  An upper limit on the nuclearite flux
may be estimated assuming that $\Phi_{max.}~=~\rho_{ DM} v/(2 \pi M)$,
 where
 $ \rho _{ DM}
\simeq 10^{-24}$ \units{g/cm^3}} represents the local DM density, and
 $M$ and $v$ are the mass and the velocity of
nuclearites, respectively.

Different indirect methods to search for nuclearites have been suggested
\cite{ruhula}.  Some exotic cosmic ray
events were interpreted as due to incident nuclearites, for example
the ``Centauro" events and
particles with anomalous charge/mass ratio [4-10].
The interpretation of those possible signals is not yet clear.
Searches for strangelets are being performed at the Brookhaven AGS \cite{ags} 
and  at the CERN-SPS \cite{na52}.\par

 Relevant direct flux upper limits for  nuclearites came from two
large area experiments using CR39 nuclear track detectors; one
experiment was performed at mountain
altitude \cite{nakamura}, the other at a depth of 10$^4$ \units{g/cm^2}
at the Ohya stone quarries \cite{orito}.

The lowest flux limits have been obtained
by examining ancient mica samples.  It should be kept in mind however
that this technique has inherent uncertainties \cite{price,ghosh}.

MACRO (Monopole, Astrophysics and Cosmic Ray Observatory) is an underground
detector located at the Gran Sasso Laboratory in Italy, at an average depth
of 3700 \units{ hg/cm^2} and at a
minimum depth of 3150 \units{ hg/cm^2}. MACRO uses three
different types of detectors: liquid scintillators, limited streamer tubes
and nuclear track detectors (CR39 and Lexan) arranged in a modular
structure of six ``supermodules" (SM's).
The overall dimensions of the apparatus are $ 76.5 \times 12 \times 9.3$ \units
{m^3}
\cite{nim93}.  The response of the three types of detectors to slow and
fast particles has been experimentally studied [18-20].
%\cite{scin,streamer,cr39}.
One of the primary aims of MACRO
is the search for superheavy GUT magnetic monopoles \cite{macromono,pub96-2}.
Some of the search methods used for this purpose,
 namely those based on the liquid scintillators and nuclear track detectors,
 may also be applied to search for nuclearites.\par

In computing the acceptances for nuclearites of the scintillator and of
the nuclear track subdetectors one has to take into consideration
the nuclearite absorption in the Earth.  In
Fig.~1 we present the fraction of solid angle from which nuclearites might
 reach
MACRO as a function of their  mass, assuming a velocity at the ground level
of $\beta = 2 \times 10^{-3}$.
For masses smaller than
 $5 \times 10^{11}$ \units{GeV/c^2}  
the nuclearites cannot reach the detector;
for $5 \times 10^{12} \leq M \leq 10^{21}$ \units{GeV/c^2} only
downward going nuclearites can reach it; for $M > 10^{22}$ \units{GeV/c^2} 
nuclearites
from all directions can reach MACRO.
The detector acceptance for an isotropic flux (at ground level) of nuclearites
has to be scaled according to such curves.

Some earlier results obtained  using
the scintillator subsystem of the lower part of the first SM
 have been published \cite{prl92}. More recent results obtained with the
 scintillator and CR39 subdetectors were
reported in Ref. \cite{icrc97}.\par
As recently suggested by Kusenko {\it et al.} \cite{qballs}, the MACRO search for
nuclearites could also apply to Supersymmetric
Electrically
Charged Solitons (Q-balls, SECS).
Q-balls are supersymmetric coherent states of squarks, sleptons and Higgs
fields, predicted by minimal supersymmetric generalizations of the Standard
Model \cite{coleman}; they could be copiously produced in the early
universe. Relic Q-balls are also candidates for cold DM \cite{cdm}.

In the following sections we present results on the search for nuclearites
in the liquid scintillator and in the nuclear track subdetectors of MACRO.
The density of the gas in the MACRO streamer tubes \cite{streamer} is too low to
allow the detection of nuclearites for the hypothesized energy loss
mechanism.

%%%%%%%%%%%%%%%%%%%%%%%%%%%%%%%%%%%%%%%%%%%%%%%%%%%%%%%%%%%%%%%%%%%%%%%

\section{Searches using the liquid scintillator
subdetector}

De R\'{u}jula and Glashow have calculated the light yield of nuclearites
traversing transparent materials on the basis of the black-body radiation
emitted along the heated track \cite{ruhula}. The light yield per unit
track length is (in natural units, $\hbar = c = 1$)
\begin{equation}
\frac{dL}{dX}=\frac{\sigma}{6 \pi^{2}\sqrt{2}}\omega^{5/2}_{max}(m/n)^{3/2}
v^{2},
\end{equation}
where $m$ is the mass of a molecule of the traversed material,
$n$ is the number of
submolecular
species in a molecule, $\omega _{max}$ is the maximum frequency for which
the material is transparent,
$\sigma$ is the nuclearite cross section and $v$ its velocity.

 In Ref.~\cite{prl92} this formula was applied
to our liquid scintillator. It was shown that the scintillator
subdetector is sensitive even to very small nuclearite masses and to low
  velocities ($\beta \simeq 5
\times 10^{-5}$). The
light yield  is
above the 90\% trigger efficiency threshold of the MACRO scintillator
slow-particle trigger system for most  nuclearite masses  
($dL/dX > 10^{-2}$ \units{MeV \ cm^{-1}}).

  The scintillators are therefore sensitive not
only to galactic
($\beta \sim 10^{-3}$) or extragalactic nuclearites
(higher velocities),
but also to those
possibly trapped in our solar system ($\beta \sim 10^{-4}$).

The nuclearite detection efficiency in the scintillator subdetector is
assumed to be similar (or larger) to that for magnetic monopoles;
the selection criteria used to search for monopole events are
also applicable for nuclearites.
No saturation effects of the detectors, electronics, or reconstruction
procedure are expected to reduce the detection efficiency of the
liquid scintillator  subdetector.

        Different monopole triggers and analysis procedures have been used
in the search for cosmic ray strangelets in different velocity domains following
the evolution of the detector.
The relevant parameters of each
actual search for nuclearites using the scintillator subdetector are
 presented in Table~1. As no candidate satisfied all the requirements,
the resulting  flux upper limits at 90\% confidence level (C.L.) are  listed
in Table~1 and presented in Fig.~2.
Some details  on the present searches  are given below.

The slow monopole trigger which includes the analog Time Over Half Maximum
     (TOHM) electronics and the digital Leaky Integrator (LI) electronics
\cite{nim93,pub96-2}
recognizes wide pulses or long trains of single photoelectrons generated
by slow particles, rejecting large and short pulses produced by muons
or radioactive decay products. 
When a  trigger occurs, the wave forms of
both the
anode and the dynode (for the 1989-91 run period) for 
each photomultiplier tube  are separately
recorded by two Wave  Form Digitizers (WFD). A  visual scan
is then performed on the selected events.
%Additional wave form analysis is performed on the selected events. 
This procedure was applied to the searches for
nuclearites with $ 10^{-5} < \beta < 3.5 \times 10^{-3}~\cite{prl92,icrc97}$.\par

The Fast Monopole Trigger (FMT) 
 is based on the time of flight  between
two layers of
scintillators. A slow coincidence between  two layers is
vetoed by a fast coincidence
between them. Additional wave form analysis is performed on the selected events.
This procedure  was applied to the search for intermediate
velocity nuclearites 
($2.5 \times 10^{-3} < \beta < 1.5 \times 10^{-2}$) \cite{prl92}.

 The  scintillator muon trigger. A fast nuclearite should produce a light
yield ($dL/dx$) at least three orders of magnitude larger than that from a
typical muon. It was
checked that no negative effects arise
on the detecting system from the larger pulse heights.
No event was found having a $dL/dx$ in both walls greater than 10 times that of
a muon.
This technique
was applied in the early analyses  for high velocity nuclearites
($ 1.5 \times 10^{-2} < \beta < 1$) \cite{prl92}.

  The Energy  Reconstruction  Processor (ERP) is a
single-counter energy threshold trigger \cite{phr1,phr2}.
The ERP analysis requires triggers in two different scintillator
planes, separated
in the vertical direction by at least 2 \units{m}, insuring a time of flight long
enough for accurate velocity measurements. The energy deposition must be
at least 600 \units{MeV} in each counter.
The ERP analysis was used to search for fast nuclearites ($\beta > 0.1$)
\cite{icrc97}.
The raw triggering efficiency for nuclearites with $\beta > 0.1$
is essentially 100\%.

  Pulse Height Recorder and Synchronous
Encoder (PHRASE)
is a system designed primarily for the detection of supernova
neutrinos \cite{phr1,phr2}. The event selection requires a coincidence between
two scintillator planes, with no more than 2 contiguous hits in each plane,
with an energy release of at least 10 \units{MeV} in each layer. It was checked that
no negative effects arise from larger pulse heights. A minimum separation of
2 \units{m} is required for hits in the two counters,
while a software cut ($\beta \leq
0.1$) is imposed in order to reject the tail of the cosmic ray muon
distribution. The particle velocity is reconstructed using the scintillator
time information. The PHRASE search for nuclearites
covers a large velocity range: $1.2 \times 10^{-3} < \beta < 10^{-1}$
\cite{icrc97}. The lower
limit corresponds to the threshold for the detection of bare 
monopoles with unit Dirac magnetic charge ($g = g_D$); 
as the light yield produced by nuclearites is larger than that
of monopoles, the nuclearite search might be extended to lower velocities.
For candidates with $\beta \leq 5 \times 10^{-3}$ we compare the duration
of the scintillation light pulse (measured by the PHRASE WFD) with the one
computed using the particle velocity; candidates with  
$5 \times 10^{-3} \leq \beta < 0.1$
are cross-checked on the basis of the measured energy loss. All the candidates
with $\beta \simeq 5 \times 10^{-3}$ are examined using both techniques, in
order to ensure the continuity of the analysis.

\section{Searches using the nuclear track subdetector}

The  nuclear track detector is located horizontally in the
middle of the lower MACRO structure, on the vertical east
wall and on the lower part of the vertical north wall. It is organised in 
modules (``wagons'') of $\sim 25 \times 25 \units{cm^2}$; a ``wagon'' contains 
three layers of CR39, three layers of lexan and 1 \units{mm} thick aluminium
absorber. Details of the
track-etch subdetector are given in Ref.~\cite{pub94-2}; the total area
is 1263 \units{m^2}.
At the point that this analysis ended we had etched
227 \units{m^2} of CR39 with an average exposure
time of 7.6 years.
In Ref. \cite {cr39} it was shown that the formation of an etchable track 
in CR39 is related to the Restricted Energy Loss (REL) which is the fraction 
of the total energy loss which remains localized
in a cylindrical region with about 10 \units{nm} diameter
around the particle trajectory \cite{benton}.
There are two contributions to REL:
the electronic energy loss ($S_e$), which represents the energy transferred
to the electrons, and the nuclear energy loss ($S_n$),
which represents the energy
transferred to the nuclei in the material.  In Ref. \cite {cr39} 
it was   
shown that $S_n$ is as effective as $S_e$ in producing etchable tracks in our 
CR39. This result was confirmed in Ref. \cite{yudong} for different 
types of CR39.
In the case of nuclearites the REL is practically equal to $S_n$; thus  Eq.~1
may be used for calculating  REL.

In Fig.~3 we present the energy loss of nuclearites in CR39;
 the calculation assumes that the
energy is transferred to the traversed material by displacing the matter
in the nuclearite path by elastic or quasi-elastic collisions. Such processes
 would produce the
breaking of the CR39 polymeric bonds, leaving etchable latent
tracks, if the energy loss is above the detector threshold.
For the MACRO CR39 the ``intrinsic" threshold is
about 20 \units{MeV\ g^{-1} cm^2}
in the condition of a chemical etching in 8N NaOH water solution at
80\degrees\units{C}; this is
shown in Fig.~3  as the lower horizontal line. The dotted line in Fig.~3
represents the REL for $g =g_D$ bare magnetic monopoles in CR39
\cite{derkaoui98}.

Several ``tracks'' were observed, mainly due to recoil protons
from  neutron
interactions or due
to polymerization inhomogeneities.
 In the conditions of the average exposure time in MACRO the number of
background tracks is about 0.5 \units{{/}m^2}  of CR39.
        When we required that the observed etch cones were
        present on at least four CR39 surfaces and were consistent with
being from the same particle track,
        all of the candidates were ruled out.

From Fig.~3 it is apparent that our CR39 is sensitive to nuclearites
of
 any  mass and with
$\beta > 1.5 \times 10^{-5}$.
 Nuclearites with mass larger than $\sim 10^{15}$ \units{GeV/c^2}
can be detected even for velocities
as small as $\beta = 10^{-5}$.
 As a consequence, the 90\% C.L. limit for $\beta \sim 1$
monopoles established
by the nuclear track subdetector ($6.8 \times 10^{-16}$ \units{
cm^{-2}s^{-1}sr^{-1}}) applies also to an isotropic flux of $ M  > 
5.6 \times 10^{22} \units{GeV/c^2}$ nuclearites.
This limit is presented in Fig.~2 as curve ``F"
and is included in Table~1.
 For lower mass nuclearites  the 90\% C.L. flux limit is twice this
value ($1.4 \times 10^{-15}$ \units{cm^{-2}s^{-1}sr^{-1}}) because
        of the solid angle effect shown in Fig.~1.

\section{Discussions and conclusions}

No nuclearite candidate was found in any of the reported searches.
The 90\% C.L. flux upper limits for  an isotropic
flux of nuclearites are presented in Fig.~2.

Because either the scintillator or the CR39 can give us a credible
nuclearite detection, we sum the independent parts of the individual
exposures
% over $\beta$-slices in which the acceptance of each particular analysis is constant 
to obtain the global limit denoted as ``MACRO'' in
Fig.~2.  This procedure ensures the 90\%\ C.L. significance of the
global limit.

All limits presented in Fig.~2 refer to the flux of nuclearites
at the level of the MACRO detector, i.e., below an average rock thickness
of 3700 \units{hg/cm^2}. To compare our limit with the limits
published by different experiments and with the limit calculated from DM
density in our galaxy,
we integrated the energy loss equation for a path
corresponding to the averaged rock thickness and for different velocities
at the detection level. Thus we obtained a relation between the nuclearite
velocities at the level of the detector and at the ground level.
Similar calculations were made for other underground experiments
\cite{orito,price}.

Fig.~4 shows the 90\% C.L. MACRO upper limit for a flux of downgoing
nuclearites
compared with the limits reported in Refs.
\cite{nakamura} (``Nakamura"), \cite{orito} (``Orito"),
the indirect mica limits \cite{price,ghosh} and with the DM limit, 
assuming a velocity
at ground level of $\beta = 2 \times 10^{-3}$. At $\beta = 2 \times 10^{-3}$
the 90\% C.L. MACRO limit for an isotropic flux of nuclearites is 
$2.7 \times 10^{-16}$ \units{cm^{-2}s^{-1}sr^{-1}}.
  In Fig.~4  we extended the MACRO limit above the
DM bound, in order to show the transition to an isotropic
flux for nuclearite masses larger than $\simeq 6 \times 10^{22}$
\units{GeV/c^2}.
\par

%%%%%%%%%%%%%%%%%%%%%%%%%%%%%%%%%%%%%%%%%%%%%%%%%%%%%%%%%%%%%%%%%%%%%
\bigskip\par
\vskip 10 pt
%\newpage
\noindent{\bf Acknowledgements.}
\vskip 0.5 truecm
We gratefully acknowledge the support of the director and of the staff of the
Laboratori Nazionali del Gran Sasso and the invaluable assistance of the
technical staff of the Institutions participating in the experiment. We thank
the Istituto Nazionale di Fisica Nucleare (INFN), the U.S. Department of
Energy and the U.S. National Science Foundation for their generous support of
the MACRO experiment. We thank INFN, ICTP (Trieste) and World Laboratory
for providing fellowships and grants for non Italian citizens.\par

%%%%%%%%%%%%%%%%%%%%%%%%%%%%%%%%%%%%%%%%%%%%%%%%5

%%%%%%%%%%%%%%%%%%%%%%%%%%%%%%%%%%%%%%%%%%%%%%%%%%%%
\newpage

\begin{center}
{\footnotesize
\vbox{\tabskip=0pt\offinterlineskip
\halign to\hsize{
\strut#& \vrule#\tabskip=.1em plus2em&
\hfil#\hfil&\vrule#&           % label
\hfil#\hfil&\vrule#&           % technique
\hfil#\hfil&\vrule#&           % #SM
\tabskip=0pt\hfil#\hfil&\hfil#\hfil&\hfil#\hfil\tabskip=.1em plus2em&\vrule#&  %beta
\tabskip=0pt\hfil#&\hfil#\hfil&#\hfil\tabskip=.1em plus2em&\vrule#&  %runs
\tabskip=0pt\hfil#&\hfil#\hfil&#\hfil\tabskip=.1em plus2em&\vrule#&  %live
\tabskip=0pt\hfil#&\hfil#\hfil&#\hfil&#\hfil\tabskip=.1em plus2em&\vrule#&  %flux
#\hfil&\vrule#\tabskip=0pt\cr
%%%
\noalign{\hrule}
\omit&height 2pt& && && && && && && && && && && &&& &\cr
&&Label&&Method&&SM's&&\multispan3\hfil$\beta$ range\hfil&&
\multispan3\hfil Run\hfil&&\multispan3\hfil Live\hfil&&
\multispan4\hfil Isotropic 90\%\hfil&&Ref.&\cr
&&in&&&&&&\multispan3&&
\multispan3\hfil period\hfil&&\multispan3\hfil time\hfil&&
\multispan4\hfil C.L. flux limit\hfil&&&\cr
&&Fig. 2&&&&&&\multispan3&&
\multispan3&&\multispan3\hfil (days)\hfil&&
\multispan4\hfil(cm${}^{-2}\,$s${}^{-1}\,$sr${}^{-1}$)\hfil&&&\cr
\noalign{\hrule}
\omit&height 2pt& && && && && && && && && && && &&& &\cr
%%%%%%%%%%%%%%%%%%%%%%%%%%%%
%PRL92
&& &&TOHM&& &&$10^{-5}$&--&$2.5 \times 10^{-3}$&& && && && &&5&.&5&$\times 10^{-15}$&&&\cr
&&A&&FMT&&1&&$2.5 \times 10^{-3}$&--&$1.5 \times 10^{-2} $&&3/89&--&4/91&&457&
.&5&&4&.&0&$\times 10^{-14}$&&\cite{prl92}&\cr
&& &&$\mu$-trigger&& &&$1.5\times10^{-2}$&--&1&& && && && &&9&.&0&$\times 10^{-15}$&& &\cr
\omit&height 2pt& && && && && && && && && && && &&& &\cr
\noalign{\hrule}
\omit&height 2pt& && && && && && && && && && && &&& &\cr
%HONG
&&B&&TOHM&&1&&$10^{-5}$&--&$3.5\times 10^{-3}$&&10/89&--&11/91&&
453&&&&5&.&6&$\times 10^{-15}$&& \cite{icrc97} &\cr
\omit&height 2pt& && && && && && && && && && && &&& &\cr
\noalign{\hrule}
\omit&height 2pt& && && && && && && && && && && &&& &\cr
%LUDLAM&ERIK
&&C&&TOHM&&1--6&& $10^{-5}$&--&$10^{-2}$&&12/92&--&6/93&&
163&.&5&&4&.&1&$\times 10^{-15}$&& \cite{icrc97}&\cr
\omit&height 2pt& && && && && && && && && && && &&& &\cr
\noalign{\hrule}
\omit&height 2pt& && && && && && && && && && && &&& &\cr
%ERP
&&D&&ERP&&1--6&& 0.1&--&1&&12/92&--&6/93&&
166&.&5&&4&.&4&$\times 10^{-15}$&& \cite{icrc97} &\cr
\omit&height 2pt& && && && && && && && && && && &&& &\cr
\noalign{\hrule}
\omit&height 2pt& && && && && && && && && && && &&& &\cr
%PHRASE
&&E&&PHRASE&& various && $1.2 \times 10^{-3}$&--&0.1&&10/89&--&12/98&&
2587&.&5&&3&.&6&$\times 10^{-16}$&& \cite{icrc97}$^*$&\cr
\omit&height 2pt& && && && && && && && && && && &&& &\cr
\noalign{\hrule}
\omit&height 2pt& && && && && && && && && && && &&& &\cr
%CR39
&&F&&CR39&&1--3&&$10^{-5}$&--&1&&9/88&--&3/99&&
-\dag&&&&6&.&8&$\times 10^{-16}$&& \cite{icrc97}$^*$&\cr
\omit&height 2pt& && && && && && && && && && && &&& &\cr
\noalign{\hrule}
}}}
\end{center}

\vskip 5 pt
{\footnotesize
$^*$ Analyses updated in this paper.

\vskip 10 pt
\dag For the CR39 it is more appropriate to quote the average exposure time of
the etched part (7.57 years).
\vskip 10 pt
}
\normalsize
Table 1. Summary of the nuclearite searches with the scintillator and with the
CR39 subdetectors. The different techniques are briefly explained
in the text. Further details are given in the quoted references.
At $\beta = 2 \times 10^{-3}$ the combined MACRO limit is $2.7 \times
10^{-16}$ \units{cm^{-2} s^{-1} sr^{-1}}. (The procedure to obtain the
MACRO limit takes care of the overlapping in beta and time ranges
of individual analyses).\\

\newpage

\begin{figure}
\begin{center}
\mbox{
\hspace{-1.6cm}
 \epsfysize=15cm
  \epsffile{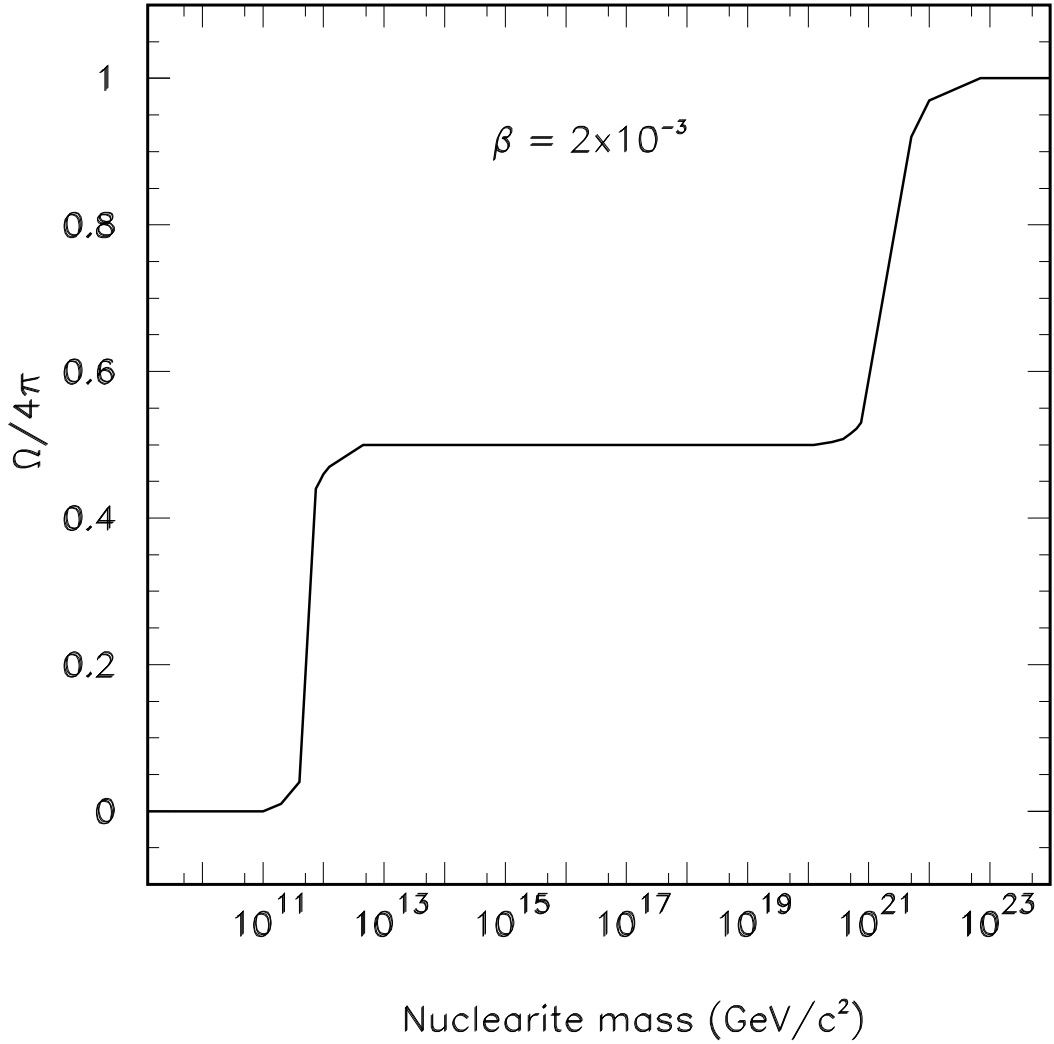}}
\caption{Fraction of the solid angle from which nuclearites with
$\beta~=~2~\times~10^{-3}$
and different masses might reach the MACRO
 detector. For masses smaller than
 $5 \times 10^{11}$ \units{GeV/c^2} the nuclearites cannot reach the detector;
for $10^{12} \leq M \leq 10^{21}$ \units{GeV/c^2} only
downward going nuclearites can reach it; for $M > 10^{22}$ \units{GeV/c^2}
nuclearites
from all directions can reach MACRO.}
\end{center}
\end{figure}

\newpage
\begin{figure}
\begin{center}
\mbox{
\hspace{-1.5cm}
 \epsfysize=15cm
  \epsffile{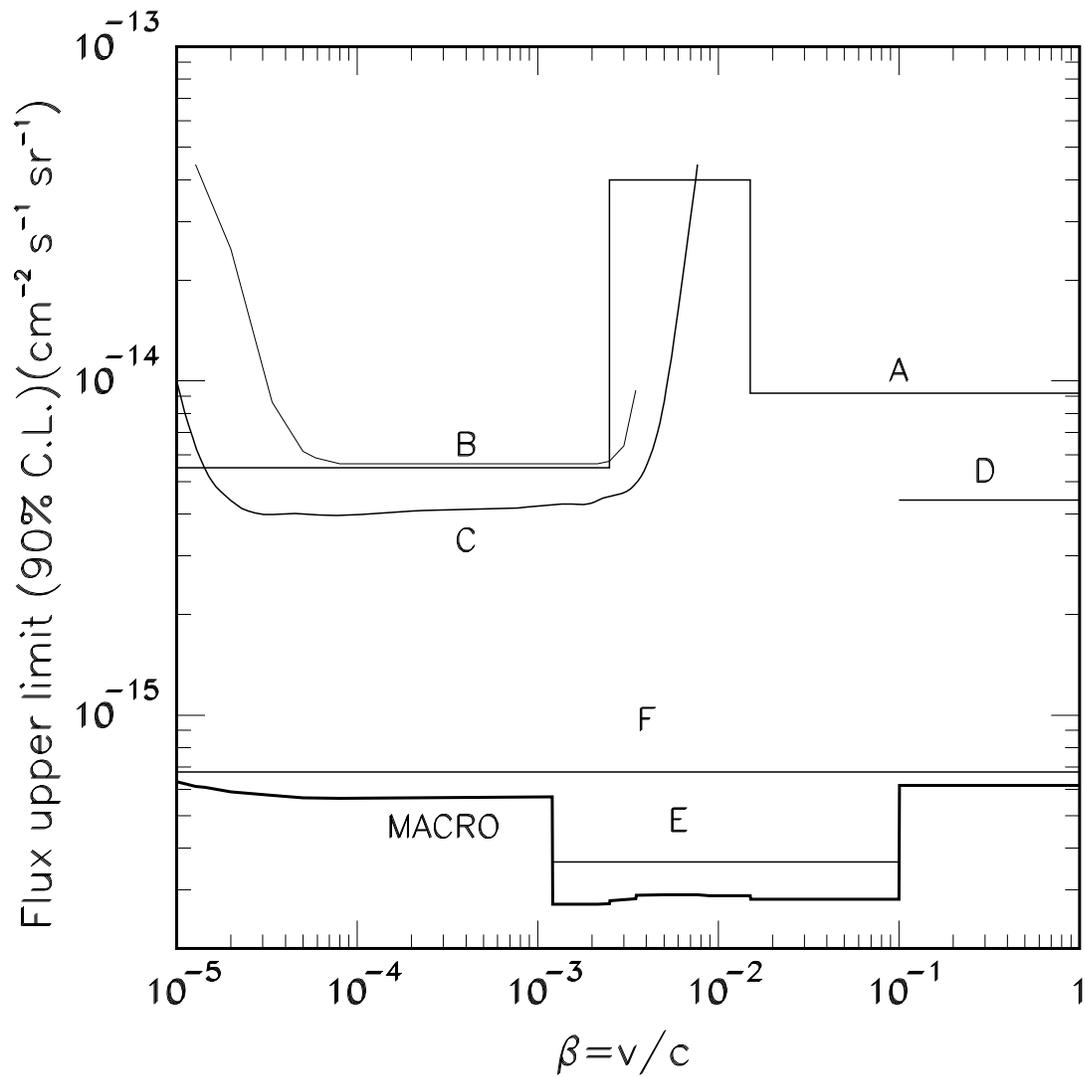}}
\caption{The $90\%$ C.L. upper limits for an isotropic flux of
nuclearites
obtained using the liquid scintillator
(curves A - E)
 and
the CR39 nuclear track
(curve F)
subdetectors;
 the bold line is the present MACRO global limit.}
\end{center}
\end{figure}

\newpage

\begin{figure}
\begin{center}
\mbox{
\hspace{-1.5cm}
 \epsfysize=15cm
  \epsffile{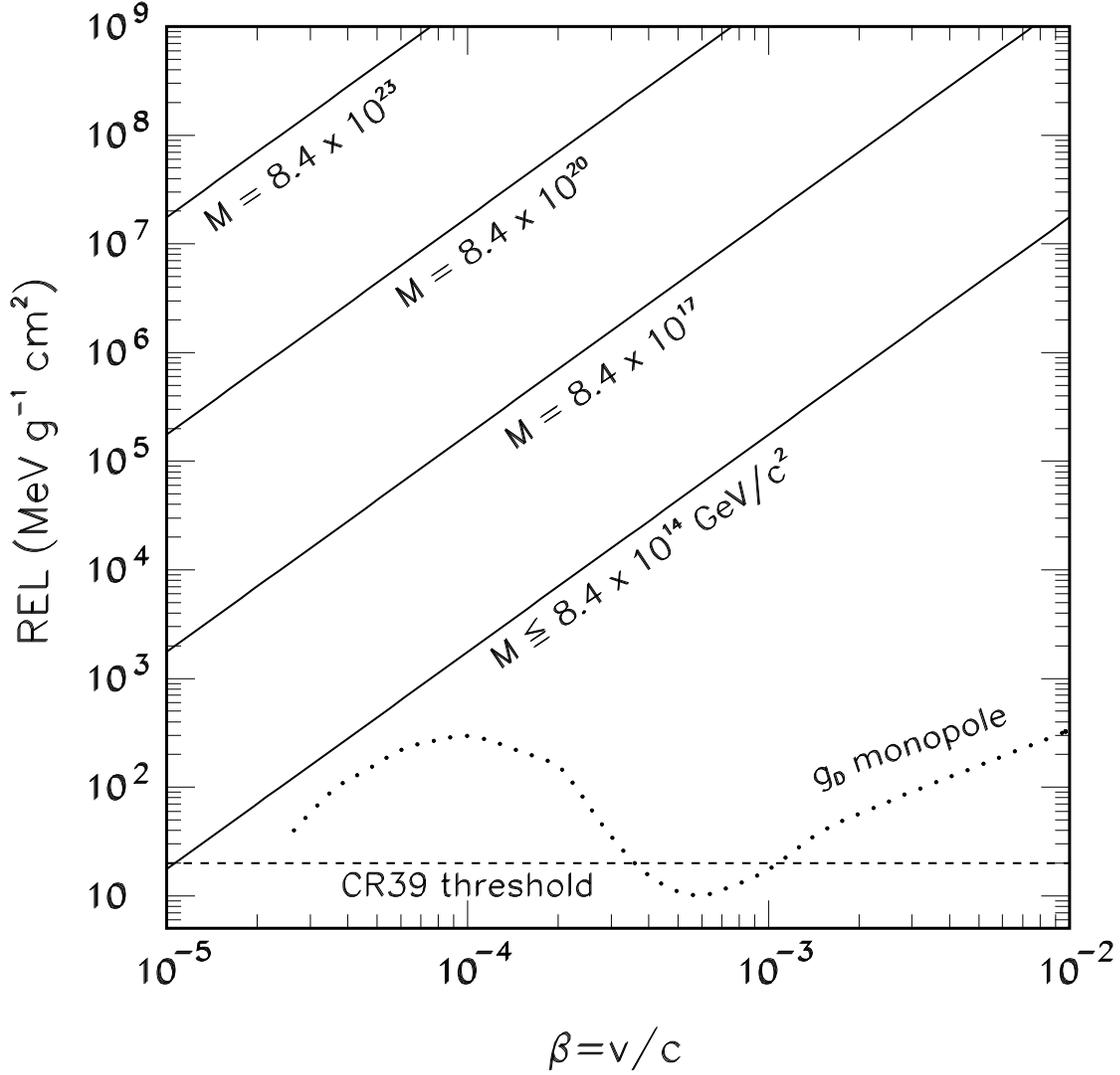}}
   \caption{Restricted
energy losses in CR39 for nuclearites of different velocities and masses
 M (in \units{GeV/c^2}).
The dotted line shows the REL of magnetic monopoles
with unit Dirac magnetic charge ($g = g_D$) in CR39
and is included for
comparison. The threshold of our CR39 is also
presented.}
\end{center}
\end{figure}

\newpage

\begin{figure}
\begin{center}
\mbox{
\hspace{-2.cm}
 \epsfysize=15cm
  \epsffile{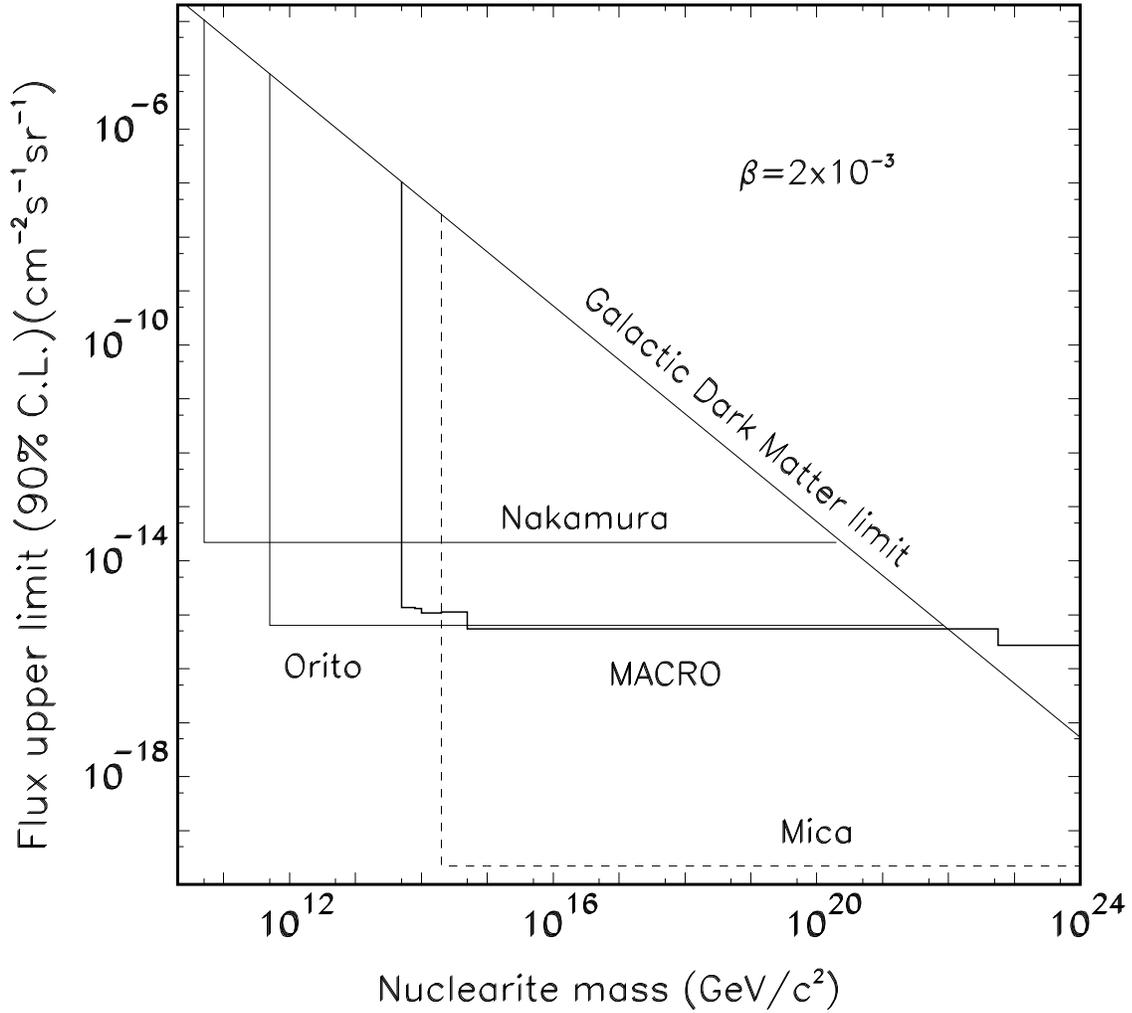}}
\caption{90\% C.L. flux upper limits versus mass for downgoing nuclearites with
$\beta = 2 \times 10^{-3}$ at ground level. Nuclearites
of such velocity could have galactic or extragalactic origin.
The MACRO direct limit is shown along with the limits from Refs.
[13] (``Nakamura''), [14] (``Orito'') and
the indirect mica limits of Refs. [15,16] (dashed line). The MACRO
limit for
nuclearite masses larger than $ 5 \times 10^{22}$\units{GeV/c^2}
has been extended and corresponds to an isotropic flux.}
\end{center}
\end{figure}

\end{document}